\documentclass[prl,twocolumn,superscriptaddress]{revtex4-2}

\usepackage{xcolor}
\usepackage{amsmath,amssymb,bm}
\usepackage{physics}
\usepackage{upgreek}
\usepackage{graphicx}
\usepackage{dcolumn}
\usepackage{dsfont}
\usepackage{nccmath}
\usepackage[colorlinks=true, allcolors=blue]{hyperref}
\usepackage{amsthm}
\usepackage{units}
\usepackage{mathrsfs}
\usepackage{hhline}
\usepackage{subfigure}

\theoremstyle{definition}
\newtheorem{problem}{Problem}
\newtheorem*{model}{Model}

\begin{document}

\title{Bespoke Pulse Design for Robust Rapid Two-Qubit Gates with Trapped Ions}

\author{Seyed Shakib Vedaie}
\email{seyedshakib.vedaie@ucalgary.ca}
\altaffiliation{These authors contributed equally.}
\affiliation{Institute for Quantum Science and Technology, University of Calgary, Calgary, Alberta, Canada T2N 1N4}

\author{Eduardo J.~P\'aez}
\email{eduardo.paez1@ucalgary.ca}
\altaffiliation{These authors contributed equally.}
\affiliation{Institute for Quantum Science and Technology, University of Calgary, Calgary, Alberta, Canada T2N 1N4}

\author{Nhung H. Nguyen}
\affiliation{Joint Quantum Institute and Department of Physics, University of Maryland, College Park, Maryland 20742, USA}

\author{Norbert M. Linke}
\affiliation{Joint Quantum Institute and Department of Physics, University of Maryland, College Park, Maryland 20742, USA}
\affiliation{Duke Quantum Center and Department of Physics, Duke University, Durham, North Carolina 27708, USA}

\author{Barry C.~Sanders}
\email{sandersb@ucalgary.ca}
\affiliation{Institute for Quantum Science and Technology, University of Calgary, Calgary, Alberta, Canada T2N 1N4}

\date{\today}
\begin{abstract}
Two-qubit gate performance is vital for scaling up ion-trap quantum computing. Optimized quantum control is needed to achieve reductions in gate duration and gate error-rate. We describe two-qubit gates with addressed Raman beams within a linear trapped-ion chain by a quantum master equation (QME). The QME incorporates the single-ion two-photon effective Rabi frequency, Autler-Townes and
vibrational Bloch-Siegert energy shifts,
off-resonant transitions,
Raman and Rayleigh scattering,
laser-power fluctuations, motional heating, cross-Kerr phonon coupling, laser spillover, asymmetric addressing beams and an imperfect initial motional ground state,
with no fitting parameters, whereas state-of-the-art methods are oblivious to these effects in the gate design procedure.
We employ global optimization to design pulse sequences for achieving a robust rapid two-qubit gate for a simulated chain of seven trapped $^{171}$Yb$^{+}$ ions
by optimizing over numerically integrated QME solutions.
Here, robust means resilient against slow drift of motional frequencies,
and rapid means gate execution where the effective Rabi frequency is comparable 
to the detuning of the laser from the ion's bare electronic transition. Our robust quantum control delivers rapid high-quality two-qubit gates in long ion chains, enabling scalable quantum computing with trapped ions.
\end{abstract}

\maketitle

\section{Introduction}
Quantum computers are implemented on various platforms---trapped 
ions~\cite{HRB08, BCMS19}, superconducting circuits~\cite{GCS17, YN03}, photonic systems~\cite{SP19}\
and neutral atoms~\cite{SBD+20}.
Although logic cycles are faster for superconducting systems compared to ion traps,
ion-trap systems currently feature better operation fidelity and qubits with longer coherence time and higher connectivity,
making them one of the leading 
platforms for achieving noisy intermediate-scale quantum (NISQ) and post-NISQ scalable 
computing~\cite{LMR+17, Pre18, Kni05}. Key performance indicators for entangling gates in trapped ions include peak power, logic-cycle time $\tau$, connectivity, and Bell-state preparation infidelity $\mathcal{I}$~\cite{BCMS19, LMR+17}, estimated from measurements of the output populations and parity-oscillation amplitudes~\cite{LBS+04, SKK+00}.

High-quality rapid two-qubit gate (2QG) control policies for preparing Bell states are vital for successful universal quantum computation, yet these gates have been achieved only in few-ion systems~\cite{CTS+21, BHL+16, GTL+16, SBK+21}. 2QGs for long ion chains operate on adiabatic time scales, where slow quantum information processing suffers from significant amounts of decoherence and fluctuations~\cite{CEN+22, PHP+22, EDN+21, WBD+19, DLF+16}.

Current control approaches for 2QGs employ certain simplifications, such as the presumption of unitary evolution, to obtain a closed mathematical expression for trapped-ion dynamics and employ greedy optimization to devise control policies~\cite{ZMD06, CDM+14, MEH+20, LLF+18, BGP+21, SIHL14, LWL+19, HM16, SSM+18, WWC+18, WWD18, SSM+20, WYW+22, Roo08}.
Therefore, such control approaches neither scale properly to fast 2QGs for chains of more than two trapped ions, nor deliver feasible control sequences. 
Thus,
we develop and validate a comprehensive model of controlled open-system dynamics described by a quantum master equation (QME).
We then numerically integrate the QME and use global optimization to devise quantum control policies for robust rapid 2QGs.
Our approach incorporates the detrimental effects of noise and decoherence directly into the gate design procedure.
We provide feasible control sequences that deliver 2QG infidelity below~0.01 for a chain of seven trapped $^{171}$Yb$^{+}$ ions in this work, limited by the specific noise environment in the setup considered. We apply Fourier analysis to motional mode dynamics and to control sequences in order to identify and interpret the underlying physics. 
Our QME method surpasses the standard approaches and delivers rapid, robust 2QG control policies that maintain high fidelity for long ion chains under realistic conditions,
thereby removing the major obstacle to implementing scalable quantum computation in trapped ions~\cite{BCMS19}.

Our aim is to reduce the gate duration~$\tau$ significantly
while respecting an in-principle lower bound due to the quantum speed limit~\cite{TEDD13}.
Reducing~$\tau$
has been explored experimentally for the `light-shift' gate employing two trapped $^{43}$Ca$^+$ ions,
achieving $\tau=1.6~\upmu\text{s}$,
$\mathcal{I}=0.002$, and $\sim$200 mW peak power~\cite{SBT+18}.
Scaling this experimental technique to several ions is very difficult due to the utilization of standing waves with precise ion indexing,
and the restriction to qubit states that exhibit linear Zeeman splitting~\cite{SIHL14}.

Our 2QG design method involves four steps.
First, we develop a comprehensive model of trapped-ion dynamics represented by a QME. Second, we validate our algorithm for computing the $4\times4$ reduced density matrix for the two target ions based on empirical data. Third, we cast the control problem of 2QG design as a feasibility problem. Finally, we employ an off-the-shelf global optimization algorithm to search for feasible robust, rapid 2QGs.

\section{Model}
Below we describe our model,
which pertains to the M{\o}lmer-S{\o}rensen gate acting on qubits manifested by
(i)~Zeeman splitting,
(ii)~hyperfine splitting
or (iii)~fine (structure) splitting~\cite{SM99, BCMS19}.

\begin{model}
A linear Paul trap with~$N$
alkali-like $\Lambda$-type~\cite{ACC+21} ions
(each labelled $\jmath\in[N]:=\{1,\ldots,N\}$)
are prepared in an electronic (meta)stable stationary state~$\ket0$;
another (meta)stable stationary state is~$\ket1$.
For a third (meta)stable stationary state~$\ket2$,
three driving fields are applied to each ion:
a beam is red-detuned by~$\Delta$
from the $\ket0\leftrightarrow\ket2$
transition,
and the other two beams
are detuned by $\Delta\pm\mu(t)$
from the $\ket1\leftrightarrow\ket2$
transition, which we call chromatic components for `red' and `blue' detuning.
Collectively, these three beams drive a bichromatic stimulated Raman transitions
for $\ket0\leftrightarrow\ket1$,
with effective Rabi frequency
$\Omega_\jmath(t)$ for the~$\jmath^\text{th}$ ion. 
The phase difference~$2\phi_\jmath$ between the two chromatic components is effectively constant~\cite{HBD+05}.
The three beams are arranged in a counterpropagating geometry to increase 
net momentum transfer along one of the principal axes of the trap with a 
wave-vector difference~$\Delta k$ and to cancel the common-mode phase fluctuation of the bichromatic fields on the spin-dependent force~\cite{HBD+05}. The time-dependent chromatic beams interact with the ions for~$t\in[0,\tau]$.
\end{model}

Ion motion is given by~$N$ collective oscillatory modes.
The rotating quadrature operator is
\begin{equation}
\label{eq:quad}
x_l(t):=a_l\exp\{-\text{i}\left((\nu_l+\delta) t+\varphi_l\right)\}+\text{hc},
l\in[N],
\end{equation}
with $a_l$ the phonon-annihilation operator and hc the Hermitian conjugate. Here
$\bm\nu=(\nu_l)\in\mathbb{R}^N$ represents motional angular 
frequencies, with $l=1$ for centre-of-mass (CoM) motion of the Wigner crystal,
next, $l=2$ is for the rocking (tilt) mode
and so on up to $l=N$.
The offset in motional-mode frequencies,
caused by slow drift in the overall trapping strength, is represented by~$\delta$.
Phase offsets are incorporated into~$\bm\varphi=(\varphi_l)\in[0,2\pi]^N$.
The coupling parameter between ion $\jmath$ and mode $l$ in the interaction is described by the Lamb-Dicke parameters~$\eta_{\jmath l}:= b_{\jmath l}\Delta 
k\sqrt{\nicefrac{\hbar}{2m(\nu_l+\delta)}}$ with~$b_{\jmath l}$ the normal-mode
coupling parameters and~$m$ the ion's mass~\cite{LBMW03}.
The $\jmath^\text{th}$ ion's position operator is
$\beta_\jmath(t):=\sum_{l\in[N]} \eta_{\jmath l}x_l(t)$.

The internal states of each ion are treated as a two-level system (2LS) with~$\ket2$ adiabatically decoupled from the Raman transition
with~$\Omega_\jmath(t)\leq\Omega_\text{max}$.
The laser detuning~$\mu(t)\equiv\mu$ is constant,
and we restrict ourselves to identical pulses for each of the two ($r$ and $s$) target ions
for the 2QG and zero driving for the other ions. 
The ion motion is not perfectly cooled to the ground state:
$\eta_{\jmath l}^2 (2\bar{n}_l + 1) \ll 1\forall \jmath,l$ with~$\bar{n}_l$ the mean phonon occupation number of mode~$l$.
Note that our condition differs from the original definition of a `warm' ion,
which presupposes that all but the motional mode used for logical operations have zero phonons~\cite{SM99}.
All motional modes are in thermal equilibrium. 
The temperature~$T$ for all motional modes is specified by mean phonon number~$\bar{n}_l$
for each mode $l$~\cite{LBMW03,TKK+00}.

\section{Quantum Master Equation}
The interaction Hamiltonian between the $\jmath^\text{th}$ ion and the electromagnetic field,
in the interaction picture with the rotating-wave approximation with respect to qubit resonance frequency,
is
\begin{equation}
\label{eq:trap_hamiltonian_2}
H_\jmath(t)
=-\hbar \Omega_\jmath(t)\cos{\left( \mu t+\phi_\jmath \right)}
    \left( \sigma^{+}_\jmath \text{e}^{\text{i}(\beta_\jmath(t)-\phi_\jmath)}
+\text{hc} \right)
\end{equation}
with ladder operators
$\sigma^+=\ket1\bra0=\sigma^{-^\dagger}$.
We use the interaction picture to avoid numerical instabilities while integrating the  QME.
The system Hamiltonian for all ions is $H(t)=\sum_{\jmath\in[N]} H_\jmath(t)$. The Autler-Townes shift
is calculated and included as a 
pulse-dependent drift of laser detuning. 
The cross-Kerr coupling  is taken into account by adding the term $\chi n_l n_{l'}$ \cite{NRJ09, MSJ03,DMHD17} to $H(t)$, where $n_l$ is the phonon-number operator of motional mode $l$ and the coupling constant $\chi$ is taken from the experimental data reported in Ref.~\cite{DMHD17}. The laser spillover onto other ions is 2\% of the Rabi frequency on neighbouring ions.

We now consider strong non-unitary processes such as heating and dephasing of motional modes, both Rayleigh and Raman photon scatterings, and laser-intensity fluctuation. The Born-Markov approximation for these dynamics~\cite{TMK+00,OIB+07,SYBH22,SM98, SM99PRA}
yields QME $\text{i}\hbar\dot{\rho}=\left[H(t),\rho\right] + \mathcal{L}[\rho]$ for~$\rho$ the state of the ions and~$\mathcal{L}[\rho]=\sum_\imath\mathcal{L}_\imath[\rho]$ the Liouvillian superoperator~\cite{GZ04, Tan06, CP17}.
Table~\ref{table:jump} presents Liouville jump terms with reasonable rate values~\cite{UBD+10, OIB+07, Lin22}. We integrate the QME to time~$\tau$.
\begin{table}[]
\centering
\setlength\extrarowheight{1.4pt}
\begin{tabular}{|c|c|c|}
    \hline
    Rate&Value [$\text{s}^{-1}$]&Jump\\
    \hline
    \hline
     $\Gamma_{l \in \{1\}}$ & 100.0 & $\sqrt{\Gamma_l} \{a_l, a^{\dagger}_l\}$  \\
     $\Gamma_{l \in \{2,\ldots,N\}}$ & 10.0 & $\sqrt{\Gamma_l} \{a_l, a^{\dagger}_l\}$  \\
     $\Gamma_{\text{MC}}$ & 27.7 & $\sqrt{\nicefrac{\Gamma_{\text{MC}}}{\pi}} a^{\dagger}_l a_l$  \\
     $\Gamma_{\text{EL}}$ & 1.5 $\times 10^{-3}$ & $\sqrt{\Gamma_{\text{EL}}} \sigma^{\text{z}}_\jmath/2$  \\
      $\Gamma_{\text R}$  & 30.0 & $\sqrt{\Gamma_{\text{R}}} \sigma^+_\jmath$ \\
       $\Gamma_{\text P}$  & 70.0 & $\sqrt{\Gamma_{\text P}} H(t)$ \\
      $\Gamma_{\text{LC}}$  & 3.0 $\times 10^{-6}$ & $\sqrt{\Gamma_{\text{LC}}} \sigma^\text{z}_\jmath$ \\
     \hline
\end{tabular}
\caption{Jump coefficients for a linear Paul trap with $^{171}$Yb$^+$ ions. Here~$\Gamma_l$ corresponds to the heating rate for the mode $l$,~$\Gamma_{\text{MC}}$ to the motional dephasinng rate per mode~$l$,~$\Gamma_{\text{EL}}$ to the Rayleigh photon scattering,~$\Gamma_{\text{R}}$ to the Raman photon scattering from the ions,~$\Gamma_{\text{LC}}$ to the laser dephasing rate, and~$\Gamma_{\text{P}}$ to the laser-intensity  fluctuations. Both~$\Gamma_{\text{EL}}$ and~$\Gamma_{\text{R}}$ are~$\Omega(t)$ dependent, and here we report the nominal values for~$\Omega = 1~\text{\nicefrac{Mrad}{s}}$.}
\label{table:jump}
\end{table}

\begin{problem}[Find a physical solution]
\label{prob:phys}
Input: number~$N$ of trapped ions,
labels
~$(r,s)$ of any pair of ions,
motional angular frequencies~$\bm\nu$,
temperature~$T$
for the CoM mode,
physical bounds for~$\Omega(t)$ and~$\mu$, $\delta_\text{tol}$
and~$\mathcal{I}_\circledcirc$.
The ions' internal states are initialized in~$\ket0$.
Output: Shortest $(r,s)$-2QG duration~$\tau$ such that $(\Omega(t),\mu)$ exists and satisfies
\begin{equation}
\label{eq:constraints_1}
    0\leq\Omega_\jmath\leq\Omega_{\text{max}}\forall\jmath,\,
    \mu_\text{min} \le \mu \le \mu_\text{max}
\end{equation}
and,
for $\ket{\Phi^{\pm}}:=\ket{00}\pm\text{i}\ket{11}$,
\begin{equation}
\label{eq:constraints_2}
 \underset{\left|\delta\right| \le \delta_\text{tol}}{\text{max}}~\mathcal{I}(\delta) \le \mathcal{I}_\circledcirc,\,
\mathcal{I}:=1-\bra{\Phi^{\pm}}\rho(\bm\Omega,\mu,\delta,\tau)\ket{\Phi^{\pm}}
\end{equation}
with~$\rho$ the QME solution.
\end{problem}

The problem is thus to reduce~$\tau$,
which is a functional of~$\Omega(t)$ and~$\mu$,
while ensuring that~$\mathcal{I}$
does not increase and the 2QG is robust.
2QG speed-up typically requires increased peak Rabi frequency~\cite{BGN+21}.
Formally, we adapt Problem~\ref{prob:phys} mathematically as a feasibility problem.
First, we discretize time by introducing a time mesh comprising~$m$ equal segments so $t\in(\tau/m)\mathbb{Z}_{m+1}$
with~$m$ a hyperparameter,
meaning a parameter that is `tuned' for optimization.
As $m= 2N+1$ suffices~\cite{CDM+14,ZMD06},
we replace~$\Omega(t)$
by~$\bm\Omega\in[0,\Omega_\text{max}]^m$.
The functional~$\tau[\Omega(t),\mu]$
is thus replaced by the function $\tau(\bm\Omega,\mu)$.

\begin{problem}[Feasibility]
Given constraints~(\ref{eq:constraints_1},\ref{eq:constraints_2}) and $\tau(\bm\Omega,\mu)\leq\tau_\text{max}$ 
for some given maximum 2QG duration~$\tau_\text{max}$,
find feasible~$\bm\Omega$ and~$\mu$.
\end{problem}

\section{Approach}
We develop a code to simulate an ion-trap system, which involves the time-dependent Hamiltonian, jump operators with time-dependent rates, input state descriptions, thermal preparation of motional modes and Bell-state preparation fidelity estimator. Subsequently, we unravel the QME employing the quantum trajectories theory~\cite{Car91} and use the C++ `Quantum trajectory class library'~\cite{SB97} to integrate the resulting stochastic quantum trajectories.
We also develop our own code for the state-of-the-art (SotA) method for pulse shaping~\cite{CDM+14,ZMD06}.

We solve the feasibility problem by employing an off-the-shelf global-optimization (GO) algorithm, specifically differential evolution~\cite{SP97,Mic17}, to search for a feasible~$\bm\Omega$ and~$\mu$ in the region~$[0,\Omega_\text{max}]^m\times[\mu_\text{min},\mu_\text{max}]$ over the uniform measure. We then compare our method to SotA.
Whereas our GO method uses the numerically integrated solutions of the QME to compute~$\mathcal{I}$, the SotA method relies on certain simplifications to obtain a closed mathematical expression for trapped-ion dynamics. 
SotA and other standard approaches then obtain a pulse sequence by minimizing the phase-space closure criterion without regard to the geometric phase condition required for the entangling 2QGs. The pulse sequence is later scaled to meet the geometric phase condition, assuming that the dynamics of motional phase-space trajectories are linear in~$\Omega(t)$.
Finally, we gain physical insight into pulse sequences and motional mode dynamics through Fourier analysis and by investigating the infidelity contributions by different noise sources.

We validate our algorithm for computing the $4\times4$ reduced density matrix for the two target ions as follows.
The predicted~$\mathcal I$ and even-parity population~$P$ must
agree with empirical results within experimental error. We choose these two quantities to validate as the experimental results along with all requisite experimental parameters and device noise characteristics are available to us~\cite{BGN+21,NLG+21}.

We simulate a chain of seven $^{171}$Yb$^+$-ions in a linear Paul trap with motional frequencies on a transverse axis spanning from 3.07~MHz (CoM) to 2.96~MHz ($7^\text{th}$ mode) and $\eta_{\jmath1}\equiv0.065\:\forall\jmath$ based on the experimental parameters in Ref.~\cite{BGN+21}. The qubit states, $\ket{F, m_F}$, are encoded into $^{171}$Yb$^+$ hyperfine states~$\ket{0,0}$ and~$\ket{1,0}$ with separation $\omega_0=2\pi\times12.642821$~GHz.
Stimulated Raman transfer occurs by virtually exciting $6^2\text{P}_{\nicefrac12}$ and $6^2\text{P}_{\nicefrac32}$.
One Raman beam is global, it
illuminates the entire crystal, 
whereas the others are a pair of tightly focused beams on the target ions.

In order to validate our algorithm, we first estimate~$\mathcal I$ for the (3,4)-2QG with~$\tau=254~\upmu$s~\cite{NLG+21}.
Our resultant~$\mathcal{I}=0.005(3)$ agrees,
within experimental error,
with the empirical result~$\mathcal{I}=0.007(4)$.
Second,
Table~\ref{tab:amfm_validation} summarizes
predicted and empirical values of~$P$ for the (4,5)-2QG over five values of~$\tau$~\cite{BGN+21}.
The data set shows good theoretical-empirical agreement, considering the uncertainties in $\eta_{\jmath l}$ and $\Omega_0$~\cite{BGN+21}.
\begin{table}[]
    \centering
    \setlength\extrarowheight{1pt}
    \begin{tabular}{|c|c|c|c|}
    \hline
     $\tau$ ($\upmu$s) & $\Omega_0/2\pi$ (\text{kHz}) & $P$ & $\bar{P}$\\
     \hline
     \hline
     190 & 286(5)& 0.977(2)& 0.978(3)\\
     200 & 182(4)& 0.995(1)& 0.991(1)\\
     250 & 149(4)& 0.994(1)& 0.991(1)\\
     300 & 113(3)& 0.991(1)& 0.991(2)\\
     350 & 107(4)& 0.991(1)& 0.991(3)\\
     \hline
    \end{tabular}
    \caption{
    We use existing data together with independently measured device noise characteristics to validate our method.
    A successful validation means reproducing the accessible empirical data within experimental error.
    The table presents empirical and estimated results for 2QGs designed using the state-of-the-art amplitude and frequency modulation method~\cite{NLG+21}. Here~$\tau$ corresponds to 2QG duration,~$\Omega_0$(error) to Rabi frequency,~$P$(error) to even-parity population and~$\bar{P}$ to estimated even-parity population.}
    \label{tab:amfm_validation}
\end{table}

\section{Results}
\begin{figure*}
    \centering
    \includegraphics[width=1.0\linewidth]{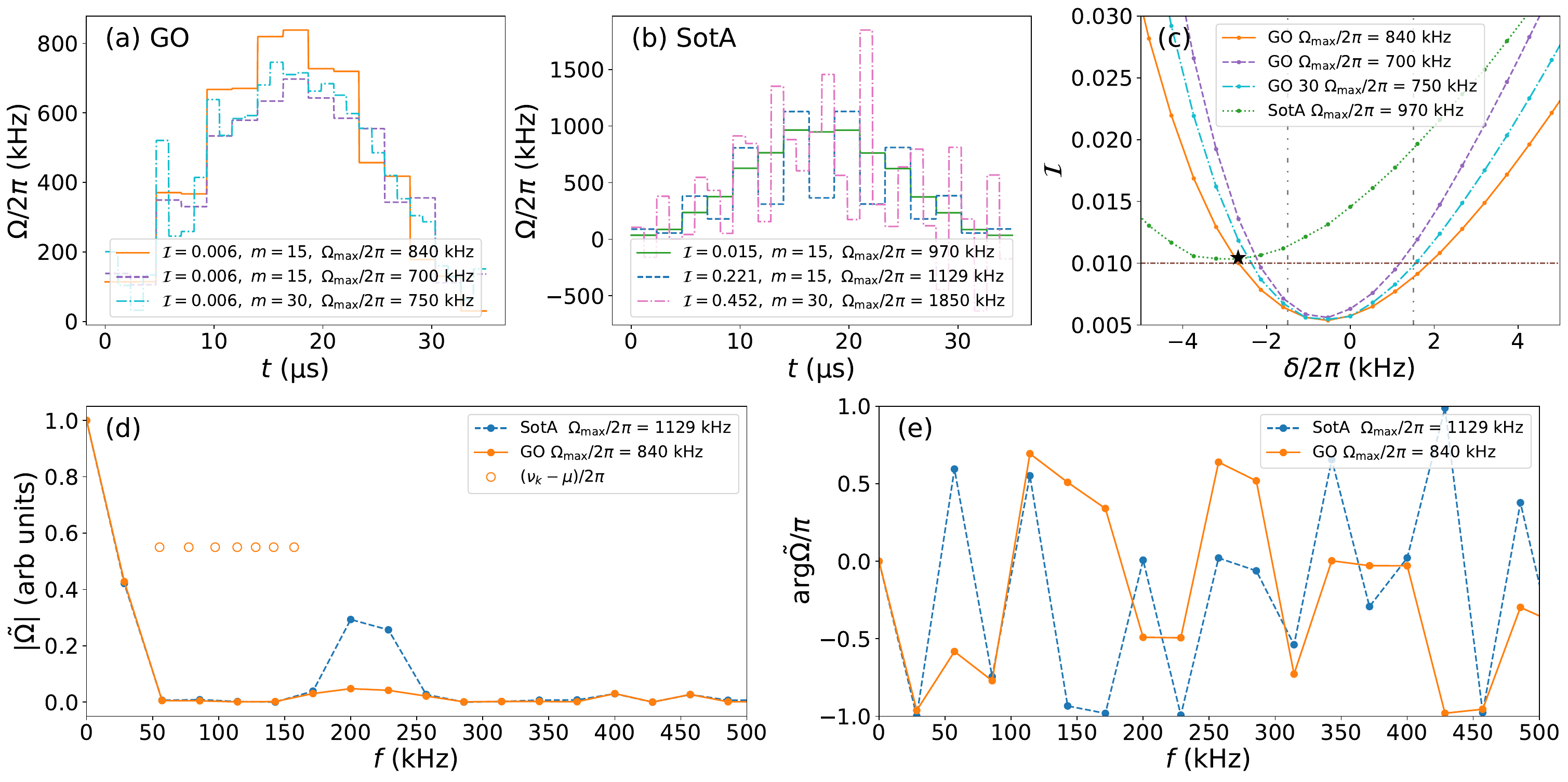}
    \caption{(3,4)-2QG pulse sequences for $N=7$, $\tau_\text{max}=35~\upmu$s,
    $\nicefrac{\mu_\text{min}}{2\pi}=2.65$~MHz,
    $\nicefrac{\mu_\text{max}}{2\pi}=3.35$~MHz,
    and~$m$ and~$\nicefrac{\Omega_\text{max}}{2\pi}$
    expressed in the legends.
    (a) Pulse sequences generated using GO with~$\mathcal{I}_\circledcirc = 0.01$. All pulse sequences have~$\mathcal{I}=0.006$.
    (b) Pulse sequences generated using SotA. The pulse sequence with~$\nicefrac{\Omega_\text{max}}{2\pi}=1129$~kHz (dashed blue) is generated at the same detuning as the GO pulse sequence with~$\nicefrac{\Omega_\text{max}}{2\pi}=840$~kHz. The other pulse sequences are generated with~$\mathcal{I}_\circledcirc = 0.02$ for~$m=15$ and~$\mathcal{I}_\circledcirc = 0.50$ for~$m=30$.
    (c) (3,4)-2QG~$\mathcal{I}$ as a function of long-term motional frequency drift~$\delta$. The horizontal line shows the feasibility condition~$\mathcal{I}_\circledcirc = 0.01$ and the two vertical lines indicate the experimentally relevant~$\nicefrac{\delta_\text{tol}}{2\pi}=1.5$~kHz domain. The SotA pulse sequence is minimized at $\nicefrac{\delta}{2\pi}=-2.67$~kHz, indicated by the~$\star$ symbol, due to not accounting for the vibrational Bloch-Siegert shift.
    Our GO method searches for 2QGs that satisfy both feasibility conditions~(\ref{eq:constraints_1},\ref{eq:constraints_2}). Therefore, the minimum of the robustness curve does not necessarily happen exactly at $\nicefrac{\delta}{2\pi}=0$.
    (d), (e) Fourier transform of $m=15$ SotA and GO pulse sequences with~$\nicefrac{\Omega_\text{max}}{2\pi}=1129$~kHz, and ~$\nicefrac{\Omega_\text{max}}{2\pi}=840$~kHz, respectively. Both pulse sequences are generated for~$\nicefrac{\mu}{2\pi}=2.89$~MHz. Circles indicate the relative frequencies of motional modes from the solutions.}
    \label{fig:pulse_shapes}
\end{figure*}

Now we show that our GO method delivers superior pulse sequences compared to SotA
in the sense that our GO method always works when SotA works, and our GO method succeeds even when SotA fails.
In Figs.~\ref{fig:pulse_shapes}(a) and~\ref{fig:pulse_shapes}(b) we present two SotA pulse sequences and three GO pulse sequences for a fast 2QG.
In this case,
all GO pulse sequences are feasible,
but neither SotA pulse sequence
is feasible unless~$\mathcal{I}_\circledcirc$ is adequately increased.

We now discuss GO and SotA performance for an increasingly fine time mesh hyperparamaterized by~$m$. We observe that, for SotA, increasing~$m$ beyond $2N+1$ increases both $\mathcal{I}$ and peak Rabi frequency $\Omega_\text{peak}:=\max{\bm\Omega}$ for fixed~$\tau$ and~$\mu$.
These increases are evident in Fig.~\ref{fig:pulse_shapes}(b). 
Specifically, no feasible SotA pulse sequence for~$m=30$ exists unless we double the~$\Omega_\text{max}$ constraint and increase~$\mathcal{I}_\circledcirc$ to 0.50.
On the contrary, the~$m=30$ GO pulse sequence is feasible under the same constraint as is the~$m=15$ GO pulse sequence.

Computing~$\mathcal{I}$ with the jump terms for individual noise sources removed, reveals the contribution of different sources of decoherence. In our case, the most detrimental are motional dephasing and laser power fluctuation, which account for 23.4\% and 22.7\% of the total~$\mathcal{I}$, respectively. This kind of analysis can help identify the most profitable avenues for hardware improvements.

Now we discuss pulse-sequence robustness against long-term drift~$\delta$ of motional frequencies.
Figure~\ref{fig:pulse_shapes}(c) presents a plot of~$\mathcal{I}$ versus~$\delta$ for the (3,4)-2QG in a seven-ion chain. The GO pulse sequences yield superior~$\mathcal{I}$ and meet the feasibility condition~$\mathcal{I}_\circledcirc=0.01$ for the experimentally relevant domain $\left|\delta\right| \le \delta_\text{tol} = 1.5$~kHz. The SotA pulse sequence does not achieve~$\mathcal{I}_\circledcirc=0.01$ and is minimized outside of the~$\delta_\text{tol} \le 1.5$~kHz domain. 
Figure~\ref{fig:pulse_shapes}(c) also indicates the greater robustness that is achieved with respect to $\delta$ by increasing, either $\Omega_\text{max}$ or $m$ for our method.
The displacement in the minimum of the robustness curve for the SotA pulse sequence is due to not accounting for the vibrational Bloch-Siegert shift in the Hamiltonian~\cite{LME08}. 

Now we Fourier analyze~$\bm\Omega$ to gain physical insight. We sample the SotA and GO pulse sequences with~$\nicefrac{\Omega_\text{max}}{2\pi}=1129$~kHz and~$\nicefrac{\Omega_\text{max}}{2\pi}=840$~kHz, respectively, at~$f_\text{s}=10$~MHz to minimize aliasing. 
The discrete Fourier transforms~$\tilde{\bm\Omega}$ of the sampled SotA and GO pulse sequences are shown in Fig.~\ref{fig:pulse_shapes}(d), where~$f\in(1/\tau)\mathbb{Z}_{(\nicefrac{f_\text{s}\tau}{2})+1}$.
The spectral width of both pulse sequences is about 28.57~kHz.
The feature around 428~kHz is an artifact due to discreteness imposed by the segment size of 2.3~$\upmu$s.
The broad bump in both spectra starting from about 142~kHz to 285~kHz corresponds to contributions about two times the pulse segment size of 2.3~$\upmu$s. The SotA pulse sequence exhibits a greater contribution in this region, which corresponds to the square-wave-like pattern of the pulse sequence with a period of twice the pulse segment size. The GO pulse sequence does not manifest such pronounced periodicity.

Phase profiles for SotA and GO are shown in Fig.~\ref{fig:pulse_shapes}(e). We note that significant differences in $\text{arg}\tilde{\Omega}$ can indicate a deficiency in the SotA pulse-sequence design method.
SotA and GO phases are similar for all frequencies (mod~$2\pi$) except between 150~kHz and 250~kHz,
where the phase of the GO pulse is antisymmetric about the midpoint 200~kHz.
This antisymmetry signifies weak coupling to the CoM mode in comparison to the SotA solution,
showing that the GO pulse succeeds because
GO avoids channelling energy to the CoM and instead concentrates the force onto the lower-frequency modes.

\begin{figure}
    \centering
    \includegraphics[width=0.85\linewidth]{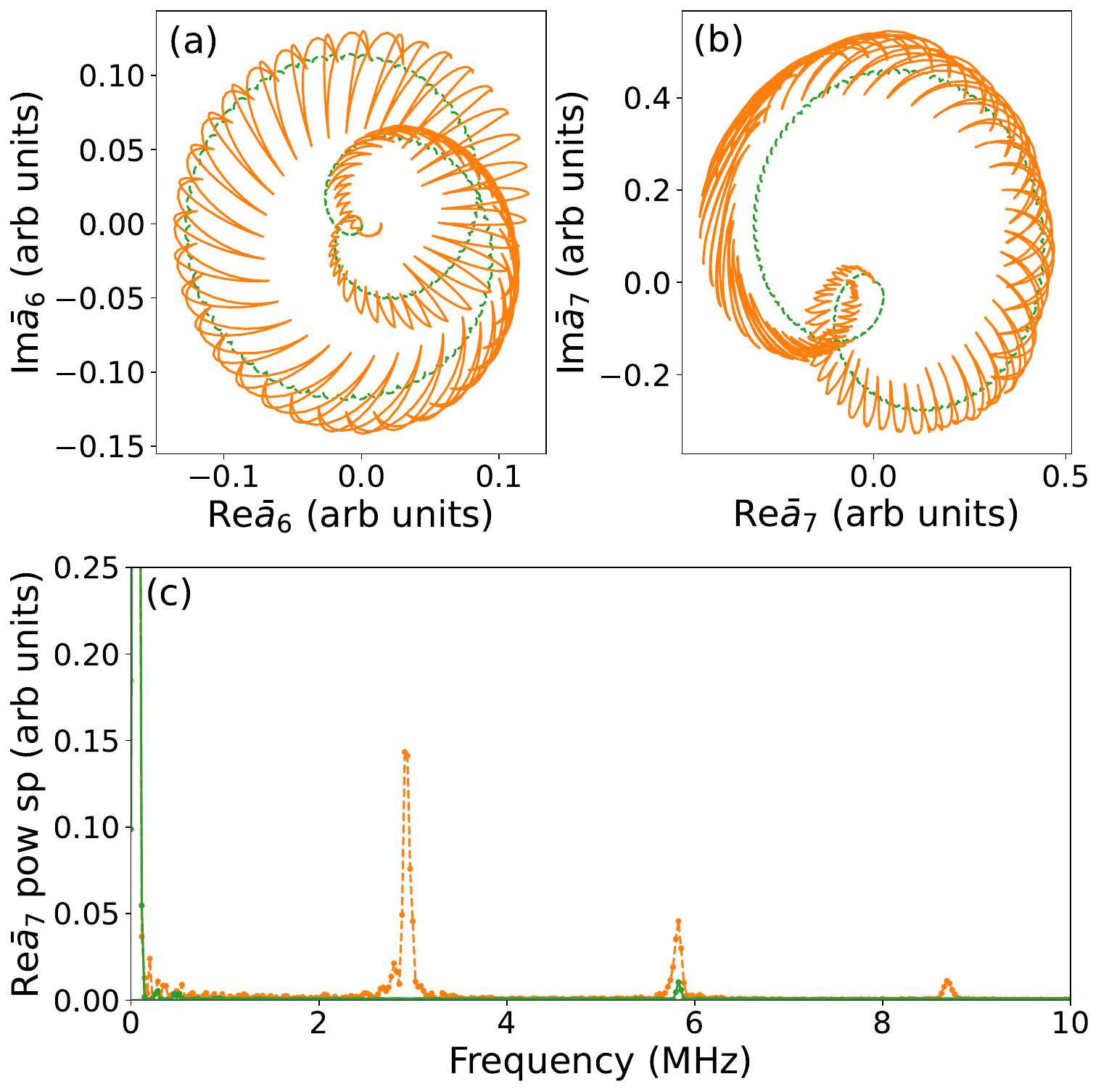}
    \caption{Evolution of phase-space trajectories for motional modes six (a) and seven (b) computed using SotA's model for trapped-ion dynamics (dashed green) and the QME based on the comprehensive model constructed in this work (solid orange).
    (c) Fourier transform of the real component of the seventh motional mode for SotA's model (solid green) and the QME (dashed orange).
    The standard approaches for 2QG gate design, including SotA, are oblivious to both the off-resonant coupling of the driving field to the bare electronic transition and the detrimental effect of the aforementioned sources of noise and decoherence.}
    \label{fig:motional_modes}
\end{figure}

Now we analyze the spectral properties of phase-space trajectories for the sixth and seventh motional modes. These trajectories are defined by the complex-valued
$\bar{a}_l(t)
:=\text{tr}(\dyad{++}{++}\otimes a_l\rho(t))$, where $\ket+ := \ket 0 + \ket 1$, and are solved for the SotA pulse sequence with~$\nicefrac{\Omega_\text{max}}{2\pi}=970$~kHz and $\nicefrac{\mu}{2\pi}=2.88$~MHz, and plotted in Figs.~\ref{fig:motional_modes}(a) and~\ref{fig:motional_modes}(b). We observe that the phase-space trajectories are jagged when solved using our QME, which is due to off-resonant coupling of the driving field to the bare electronic transition. The SotA model is oblivious to this effect. Therefore, SotA pulse sequences fail to close the phase-space trajectories into orbits at the end of 2QG implementation, which leads to residual entanglement between the electrons and the motional modes.

The Fourier transform of the real component of the seventh motional mode is shown in Fig.~\ref{fig:motional_modes}(c).
The QME approach shows frequency contributions around multiples of $\mu$, which correspond to the higher-order effects of the off-resonant bare electronic transitions.

\section{Conclusion}
Although we provide a study of our 2QG design method on a simulated chain of seven $^{171}$Yb$^+$-ions in a linear Paul trap, our method is readily extensible to longer chains of ions, gate duration and physical constraints. We validate our model by computing~$\mathcal{I}$ for two of the state-of-the-art 2QG methods within experimental error. Thus, we show that the physics we extract applies across other control scenarios, error environments, and system sizes considered within the scope of this work.
The maximum dimension of the Hilbert space we use for simulating the chain of seven $^{171}$Yb$^+$-ions is 320000, which includes the Hilbert space for two ions plus two more neighbouring ions due to laser spill-over. Additionally, our optimized implementation of the QME integrator dynamically lowers or increases the size of Fock space for motional modes based on their occupation probability and adjacency to~$\mu$.
The parallelized implementation of our GO method takes about~45 min on a computer with~40 cores to obtain a feasible 2QG solution, which is well within the time frame required for an experimental test. We expect that with further optimization this time can be considerably reduced as well.

Our work uncovers principles for 2QG design that are informed by a comprehensive model of trapped ions. To minimize off-resonant carrier modulation, the pulse sequence should begin and end smoothly at zero strength since the amplitude of the modulation depends linearly on the Rabi frequency. Moreover, the pulse symmetry, together with the appropriate detuning, contribute to close the phase-space trajectories and improve the robustness.

\section{Acknowledgements}
We acknowledge the traditional owners of the land on which this work was undertaken at the University of Calgary: the Treaty 7 First Nations (www.treaty7.org). SSV would like to thank MITACS. BCS and EJP acknowledge the support of the Major Innovation Fund, Government of Alberta, Canada. 
NML acknowledges support from the National Science Foundation (QLCI grant OMA-2120757) and the Maryland—Army-Research-Lab Quantum Partnership (W911NF1920181). Support is also acknowledged from the U.S.~Department of Energy, Office of Science, National Quantum Information Science Research Centers, Quantum Systems Accelerator. We thank Alaina M.~Green for useful discussion on the simulation of motional coherence time. The computational work is enabled by the support of Compute Ontario (www.computeontario.ca) and Calcul Qu\'ebec (www.calculquebec.ca).

The data that support the findings of this study are available from the corresponding authors, SSV and EJP, upon reasonable request.

SSV and EJP contributed equally to this work.

\end{document}